# Automatic quality control framework for more reliable integration of machine learning-based image segmentation into medical workflows

Elena Williams[1,2,8], Sebastian Niehaus[1,3,4,8], Janis Reinelt[1], Alberto Merola [1], Paul Glad Mihai[1], Kersten Villringer[5], Konstantin Thierbach[1], Evelyn Medawar[1,*], Daniel Lichterfeld[1], Ingo Roeder[3,6], Nico Scherf[3,4,7], Maria del C. Valdés Hernández[2]

1 - AICURA medical, Bessemerstrasse 22, 12103 Berlin, Germany
2 - Centre for Clinical Brain Sciences, University of Edinburgh
3 - Institute for Medical Informatics and Biometry, Carl Gustav Carus Faculty of Medicine, Technische Universität Dresden, Fetscherstrasse 74, 01307 Dresden, Germany
4 - Max Planck Institute for Human Cognitive and Brain Sciences, Stephanstrasse 1a, 04103 Leipzig, Germany
5 - Center for Stroke Research Berlin, Charité, Universitätsmedizin Berlin, Berlin, Germany
6 - National Center of Tumor Diseases (NCT) Partner Site Dresden, Fetscherstrasse 74, 01307 Dresden, Germany
7 - Center for Scalable Data Analytics and Artificial Intelligence ScaDS.AI, Dresden/Leipzig, Germany, 04103 Leipzig, Germany
8 - These authors contributed equally: Elena Williams and Sebastian Niehaus.
*Contact: evelyn.medawar@aicura-medical.com

## Abstract

Machine learning algorithms underpin modern computer aided diagnosis software, which has proved valuable in clinical practice, particularly in radiology. However, inaccuracies, mainly due to the limited availability of clinical samples for training these algorithms, hamper their wider applicability, acceptance, and recognition amongst clinicians. We present a framework to evaluate state-of-the-art automatic quality control (QC) methods that can be implemented within these algorithms to estimate the certainty of their outputs and to sort out quantitatively insufficient predictions based on established thresholds. We demonstrate that such a framework reduces the number of segmentations which need manual evaluation by 84.5%. We validated the framework with two different QC approaches on a brain image segmentation task identifying white matter hyperintensities (WMH) which are particularly challenging to segment due to their varied size, and distributional patterns in magnetic resonance imaging data. Our work reveals how the evaluation framework for machine learning algorithms can help to detect failed segmentation cases and guide in the selection of a QC method, thereby making machine learning-based automatic segmentation more reliable and suitable for clinical practice.

## Introduction

Medical image segmentation is essential for clinical decision-making, diagnostics and treatment planning. However, manual assessment of medical images, incorporated in nationwide medical-support systems, is time-consuming and can lead to inaccurate results impacting clinical decisions. Despite Convolutional Neural Networks (CNN) being able to already perform image segmentation tasks with accuracies equaling or surpassing those of humans (1), inaccuracies cannot be completely avoided, which can lead to reduced acceptance by clinicians. Thus, it is important to understand the constraints and instabilities of CNN models and to assess the quality of the reported results (e.g., identify failed segmentation cases where the predicted output was wrong or inaccurate) making the models safer and more reliable. As manual quality control on a large scale is not attainable, automated methods are being developed. In medical image segmentation, the most common quality control (QC) approaches employed to detect failed results include i) estimation and visualisation of uncertainty and errors (2–9), ii) aggregation of the uncertainty and error estimates into a single score (2,3,7,10–14), and iii) prediction of quality class or Dice coefficient, a measure commonly used to assess the spatial agreement between the ground truth and the predicted segmentation map (7,15–20).

The architecture of standard CNN models does not allow uncertainty estimation for predicted outputs (21). Although efforts have been done recently to incorporate uncertainty in deep learning schemes (22–24), the application of Monte Carlo (MC) dropout in the hidden layers of the neural network (5–7,9,14,15,21,25) continues being the most widely used. To extract the uncertainty, the model with MC dropout layers samples N predictions for every input. The final prediction is estimated by averaging the N prediction maps. Then, using various measures such as entropy and predictive variance (3–5,14,15,26), the voxel-wise uncertainty levels are extracted from the final prediction map. In contrast, error estimates are obtained with the application of generative models which aim to approximate the model distribution to the true data distribution to generate new samples with some variations (16,19). Error maps are usually calculated by subtracting voxels of generated images or segmentation maps from the voxels of original input samples. It is assumed that if the segmentation output is not optimal, an obvious difference between the generated sample and the input sample would be apparent (19,27).

Uncertainty and error maps can be used for visual assessment or correction of segmentation results (2,12). They can also be used to train a regression model to predict Dice coefficients (7,15,20) or aggregated into a single score that could potentially substitute true Dice values (14). The score can comprise measures like the intersection over union overlap (14), quality metric (12,13), mean voxel-wise uncertainty (7,14) and the pixel-wise sum (10,11). The performance of predicted Dice and an aggregated score in detecting failed segmentation results is oftentimes evaluated by estimating the Pearson correlation coefficient or drawing scatterplots with the true Dice values (15,20,27). It is assumed that the more linear the relationship, the more likely the predictive or aggregated measure will successfully substitute Dice values. In addition to correlation, the validation can be performed by calculating the mean squared error between the predicted and true Dice values (12,16). The problem with the current evaluation methodology is that it only evaluates the outcome of the QC method rather than the overall performance of the failure detection system. Such research is not purpose-driven and does not allow a proper comparison between the QC methods. We propose an evaluation framework for QC assessment and method selection. The framework includes a magnetic resonance imaging (MRI) filter which is based on a pre-defined success threshold. This allows the application of common performance metrics such as i) Precision and Recall, ii) an estimation of how many failed segmentations were identified and iii) the mean Dice value after these segmentations were removed from the data set. The presented approach provides deeper insights into the QC performance.

## Material and Methods

We combined data from four databases of brain magnetic resonance images (MRI) from patients with white matter hyperintensities (WMH) of cardiovascular and Multiple Sclerosis (MS) origins, as they have a similar appearance (29). By combining data from different sources, we ensured variations in the training sample to overcome overfitting problems common when a single imaging dataset is used [36]. The datasets used in this project were:

1. The dataset from the MICCAI WMH Challenge (MICCAI17) (31).
2. Data from the MS Lesion Segmentation Challenge at ISBI 2015 (MS15) collected from five patients at different time points (32). Only data from one time point was used in the experiments to prevent overfitting.
3. The dataset from MICCAI MS segmentation challenge 2016 (MS16) (33) .
4. The MR image database of MS pathology (MS17) (34).

Details of these datasets are shown in Table 3. In our experiments we used the ground truth segmentation maps as well as the T1-weighted and fluid attenuation inversion recovery (FLAIR) MRI sequences, which show different contrasts between tissue types. The T1-weighted MRI sequence offers a good contrast between the healthy brain tissues (i.e., normal-appearing white and grey matter) while FLAIR helps to distinguish pathologies present in white matter (35). In total, there were T1-weighted and FLAIR MRI data samples from 105 patients. All experiments were run using a 5-fold cross-validation. The data samples were randomly assigned to each fold and split into training, validation, and test sets with a size of 68, 16, and 21, respectively. The details on data preprocessing are presented in **SI**.

[Table 2]

*Evaluation Framework*

Using the results of a literature review we selected automatic QC approaches which are the most suitable for a CNN-based WMH segmentation task (see further details in **SI "Literature Review"**). We implemented and evaluated the performance of selected QC methods such as aggregation of uncertainty and error estimates into a single score and prediction of Dice coefficients.

To validate the performance of QC approaches, we used the evaluation framework in which we assess several parameters: Pearson correlation coefficient, mean squared error (MSE), Precision, Recall, number of detected failed cases and mean Dice after the detected failed segmentations were removed from the test data set. We applied an MRI filter in which a defined threshold quality value is used to automatically remove failed samples. For example, we considered the prediction maps with a Dice value of less than 0.75 to be poor quality and using this threshold we converted the sets of true Dice values and predicted Dice values into Boolean vectors. This allowed us to measure Precision and Recall (see further details in **SI**). We also estimated the number of samples removed and compared the mean Dice before and after removal.

*Uncertainty and error estimation*

To implement a segmentation model, we chose the three-dimensional (3D) U-net architecture proposed by Kayalibay et al. (28) as it allows combining multiple segmentation maps on different scales. As **Figure 4 (a)** shows, to allow uncertainty estimation, MC dropout layers were added after every convolutional block with a rate of 0.2 as in (14,21,36). To train the MC U-net we used a combination of T1-weighted and FLAIR volumes. For the gradient descent optimisation, we used the Dice loss function, where a minimum value of 0 indicates a perfect overlap between the ground truth and predicted segmentation map. Further training parameters and data preprocessing details are described in the **SI "Reconstruction U-net training parameters"**.

To obtain final prediction maps and extract the uncertainty, we passed forward the MC U-net model 20 times for every input; $N = 20$ was chosen as it was shown to capture well the uncertainty levels (6,15). To estimate uncertainty, we chose an entropy measure that previously has shown to capture well the modelling uncertainty for MS lesion segmentation outputs (4). The entropy measure allowed us to assess the amount of information inherent to the model's predictive density function at every voxel.

To estimate the error maps, we implemented a reconstruction (Rec) U-net as in (16). We decided not to use conditional GANs (cGANs), because these require a multi-class segmentation map (12,19,27), whereas in the case of WMH segmentation only binary maps are used. As shown in **Figure 4 (b)**, to create a reconstruction U-net we changed the last layer of the baseline U-net architecture with a ReLU activation function (37). We used the FLAIR modality as input due to its characteristic of highlighting the white matter pathology. Similar as (16), for the training we inserted zeroes in 3D FLAIR images in the areas where the ground truth WMH segmentation regions were located for the network to learn recovering the missed areas. To train the reconstruction model, we used the structural similarity index measure (SSIM). Further implementation details are given in **SI "CNN Regression training parameters"**.

[Figure 4]

*Voxel-wise sum aggregation*

Similar as in (10,11), to aggregate uncertainty and error maps we used a voxel-wise sum (VS) measure. As demonstrated in (4,29), small isolated WMH clusters are more difficult to estimate correctly. Therefore, it is often the case that WMH are poorly segmented in brain images with only small punctate clusters, manifesting in lower Dice values. Considering that uncertainty and error estimates are often based around the WMH regions (20,27), we would expect that patient image data with a small WMH load would have lower VS coefficients. Consequently, our hypothesis is that small VS values would correlate with low Dice values. The more correlated the values, the more likely the proposed method can be used to estimate the quality in situations where the ground truth is not available and real Dice values cannot be obtained.

The minimum and maximum VS value depends on the voxel value range and the size of the volume: the bigger the volume size, the more voxels there will be in the uncertainty and error maps and the higher the VS can get. $Eq.1$ shows an example of how the $VS_i$ values were calculated using uncertainty maps $\widetilde{H}_{\hat{y}_{i,avg},v}$, where $v$ is a voxel:

$$VS_i = \sum_{V=0}^{V} \widetilde{H}_{\hat{y}_{i,avg},v} \qquad Eq. 1$$

*Dice Regression*

As per the literature review, another common QC approach is to train a regression model to predict Dice values. To evaluate this approach, we built a 3D CNN Regression model, namely Dice Regression. The model architecture is shown in **Figure 4 (c)**.

To compare which features are better at predicting the quality, we carried out three experiments using the following inputs:

1. Image-prediction pair.
2. Uncertainty-prediction pair.
3. Error-prediction pair.

For every data sample in the training, validation, and test sets across the 5-folds, we saved the average prediction maps estimated using i) an MC U-net, ii) its respective Dice scores, iii) the uncertainty maps, and iv) error maps. For the estimation of the error maps in this experiment, instead of the ground truth segmentation maps, we used average prediction maps to place 0s into the original FLAIR images. To optimise the Adam gradient descent algorithm, we used the Huber loss function. Further training parameters are given in the **SI**.

The analyses were performed using Python (38). The network architecture and modelling process were implemented using TensorFlow (39). The training was run on a CPU-GPU hybrid system utilising 2 NVIDIA Tesla V100 GPUs (16 GB each) to parallelise the computation.

## Results

In this work, we developed a framework and validated two state-of-the-art QC approaches within this framework for CNN-based segmentation models: the aggregation of uncertainty and error estimates into a single score and the prediction of Dice values. The QC evaluation framework was evaluated in one of the most widely researched tasks in medical image analysis: the segmentation of white matter hyperintensities (WMH) from structural brain scans using magnetic resonance imaging (MRI). We used MRI data from multiple publicly available data sets acquired across different sites. In **Figure 1** we compare manual evaluation of automatic segmentation to the proposed automated evaluation framework.

[Figure 1]

CNN-based segmentation models such as the U-net (28) are already used in clinical workflows, however, each prediction requires manual validation by medical professionals (**Figure 1a**). As demonstrated in **Figure 1c**, failed prediction maps are oftentimes undersegmented and require manual delineation. An application of the QC method and MRI filter optimises detection of failed segmentations and requires manual review of only a fraction of segmentation maps (**Figure 1b**). Based on average number of failed cases removed using the most promising QC methods, such as uncertainty VS score and predicted Dice score using uncertainty-prediction and error-prediction pair, we estimated that 15.5% out of 21 segmentation maps need manual validation, meaning 84.5% of MRIs are successfully segmented (**SI Figure 1**). The majority of the MRI scans, around 84.5%, are successfully segmented and can be directly sent to the DICOM server overpassing the manual validation.

In addition to the correlation coefficient and mean absolute error metrics, the evaluation framework includes an application of an MRI filter which allows estimation of common performance measures such as Precision and Recall, number of detected failed cases and mean Dice score after the detected failed cases were removed. To estimate uncertainty, we implemented the U-net and added MC dropout in the hidden layers of the model. The uncertainty maps were obtained with an application of an entropy measure on the predicted segmentation maps. To obtain error estimates, the U-net was adjusted to perform a reconstruction task, and the error maps were estimated by subtracting the reconstructed MRI images from the real ones. We visualised the uncertainty and error maps to compare them with the ones from the previous findings. The aggregation was performed using a voxel-wise sum operation (VS) over uncertainty and error maps. To predict the Dice values, we implemented a Dice Regression model and ran experiments with various inputs.

*Visualization of uncertainty and error estimates*

Examples of estimated uncertainty and error maps for brains with different WMH burdens are shown in **Figure 2**. Consistent with findings from state-of-the-art methods applied to the same task in different cohorts (reviewed in (29)), sometimes our models undersegmented small WMH (**Figure 2a**, Prediction Map) as well as large WMH (**Figure 2c**), while showing high accuracy segmenting WMH in brains with medium WMH load (**Figure 2b**). Uncertainty was mostly located on the borders of the WMH regions and larger differences were found more frequently on the error maps for brains with large WMH burden (**Figure 2c**). There were cases where artifacts were labeled as WMH in the ground truth maps. Such cases resulted in high uncertainty levels (**Figure 2c**, Uncertainty Map), yet were correctly classified as WMH in the prediction map. Although the reconstructed images look almost identical to the true images, there are prediction errors, particularly in the WMH regions, displayed by the error maps as intensity differences.

[Figure 2]

*Aggregation of uncertainty and error estimates*

The scatterplots in **Figure 3 (a-b)** demonstrate that the values of the VS extracted from the uncertainty maps and the respective Dice coefficients are relatively linearly distributed in contrast to the aggregated error estimates, which exhibit larger confidence intervals.

[Figure 3]

The VS performance results across 5-folds are summarised in **Table 1**. The mean Pearson correlation coefficient between the uncertainty VS and Dice score was 0.52 ($P < 0.05$). No significant correlation was observed between Dice score and error VS values ($R$ -0.25, $P > 0.1$).

[Table 1]

We removed the samples with an uncertainty VS lower than 1100 as these instances correlated with low Dice coefficients (**Figure 3a**). After the removal of failed segmentation maps, the mean Dice value increased from 0.84 to 0.85 by detecting 3.2 failed cases on average with Precision of 0.93 and Recall of 0.89. Using a threshold error VS value of -100,000, which correlated with the Dice value of one of the observations with lowest Dice score (**Figure 3b**), we observe a slight drop in mean Dice value from 0.84 to 0.837, with a Precision of 0.48 and a Recall of 0.83. The performance per fold is shown in **SI Figures 4-5**.

*Dice Regression*

**Figure 3** shows three scatterplots (c-e) with predicted and true Dice values from one of the folds. Dice values predicted with image-prediction and uncertainty-prediction pairs were higher than the true Dice values. However, Dice values predicted with error-prediction pairs were slightly lower than the true Dice values. The observations on the scatterplots (**Figure 3c**, **Figure 3d**) were located closer to the regression line and had narrower confidence intervals.

**Table 1** shows that the Dice predictions with uncertainty-prediction input produced the lowest mean absolute error (MAE 0.0486) and highest correlation coefficient (r 0.71, $P < 0.0001$) showing a strong and positive relationship between the predicted and true Dice. The highest MAE was achieved with the error-prediction input and the lowest correlation coefficient with the image-prediction pair. Dice values predicted with an input of error-prediction pairs allowed us to identify the highest number of failed segmentations (**Figure 3e**) and produced the highest increase in the median Dice value, namely from 0.84 to 0.87, with median Precision of 0.941 and Recall of 0.833. The results per fold are shown in **SI Figures 4-5**.

## Discussion

The safety of the application of machine learning models is crucial for its successful adoption in clinical practice. In this study, we developed a framework to evaluate state-of-the-art QC methods which aim at detecting failed CNN-based WMH segmentation results. In addition to estimating correlation coefficients and mean absolute error, we applied an MRI filter to calculate Precision and Recall, the number of detected failed cases and mean Dice after removing these segmentations which allowed us to successfully compare the performance of the QC methods. We showed how the QC evaluation framework can reduce the amount of time spent by clinicians on manual validation by reviewing on average 84.5% less of segmentation results, making automatic segmentation more reliable and suitable for clinical routines. The most popular QC approaches are aggregation of uncertainty and error estimates into a single score, and prediction of Dice coefficients. Among these, Dice prediction yielded the most encouraging results for automatic QC. From our experiments, we saw the highest number of detected failed cases and an improved mean Dice value using a model trained with a combination of error and prediction maps.

Similar to what is reported in (10,11), we found a significant correlation between the VS values extracted from uncertainty maps and Dice coefficients. We also saw an improved mean Dice value after we removed failed segmentation maps based on these VS values as well as high Precision and Recall. This supports our hypothesis that VS values calculated based on uncertainty maps can be used as a substitute for Dice coefficients in cases when the ground truth is not available. We assumed the same would be true with aggregated error maps. However, in this case the VS measure did not show any significant correlation with Dice values, there was a slight decrease in Dice after removing failed cases and we had to reject our null hypothesis.

In the experiments with Dice prediction, we achieved a similar performance to the one demonstrated in (15,16,20,27). Unlike the other works, we validated not only the model performance but also the efficacy of Dice prediction in failure detection. We ran the experiments using three input types to see which features would work best in identifying failed segmentation results. We note that, although an input of uncertainty and prediction maps showed the highest correlation and the lowest error with true Dice values, it was moderately effective for automatic QC which highlights the importance of the proposed evaluation framework. The best QC performance was demonstrated using Dice coefficients predicted with an error-prediction pair. This can be attributed to the fact that the model trained with an input of error-prediction pair slightly under-predicted the Dice values whereas the models trained with image-prediction and uncertainty-prediction pairs slightly over-predicted the Dice values.

The main strength of this study is the demonstration of how the QC evaluation framework can optimise detection and correction of inaccurate segmentations in biomedical imaging workflows. For every QC approach we identified unsuccessful segmentation maps and evaluated it using 5-fold cross-validation, based on which we showed how successful the method was in identifying failed cases. None of the previously published studies i) compared mean Dice values before and after detection of unsuccessful segmentation results, ii) analysed the number of failed cases identified and iii) estimated Precision and Recall coefficients to show how well the proposed QC approach performed in practice. Usually only the error rate or correlation coefficient between the predicted and true measure of segmentation is being reported. This information is not sufficient to evaluate the efficacy of an automatic QC system. An application of the proposed evaluation framework would increase so needed trust in the usage of digital technologies in clinical context (30). From the literature review performed, we noted that automatic QC for medical image segmentation has become an actively developing field of research in the last three years. However, very few papers so far attempted to summarise the latest findings and evaluate the performance of state-of-the-art QC methods in practice (2,15,20).

Although the QC evaluation framework generalises well and can be applied to various types of segmentation tasks, there are limitations related to the selection of QC methodology and input data which would require some adaptation to the design presented in this work. For example, given the variability inherent to the location and size of WMH, a different selection of uncertainty measures or aggregation scores would be needed. Moreover, the threshold for sorting out the successful segmentations from unsuccessful ones would need to be adjusted for a specific use case. Further validation of QC methods using a different type of imaging data and segmentation problems such as cardiac MRI or liver CT scans is therefore needed.

## Conclusions

The unique contribution of this work is the proposal of an evaluation framework and its assessment of state-of-the-art automatic QC methods. We apply the QC methods to a challenging and clinically relevant brain imaging segmentation task and demonstrate the advantages of using the QC validation framework. By providing an overview of the performance of these QC methods in a highly heterogeneous sample in terms of image acquisition protocols and patients' health conditions, this study intends to assist researchers in developing robust medical image segmentation pipelines by guiding the selection of the QC strategy. We illustrate the necessity of the use of QC methods as a key step for the optimal interaction between physicians and machine learning systems.

## Acknowledgments

### Funding
This work was supported by Row Fogo Charitable Trust (grant no. BRO-D.FID3668413, MVH).

### Author contributions
Conceptualization: EW, SN, JR
Methodology: EW, SN, AM, DL
Investigation: EW, SN, PGM, KV
Visualization: EW, SN
Supervision: NS, MVH, IR
Writing—original draft: EW, SN
Writing—review & editing: EW, SN, MVH, JR, AM, PGM, NS, IR, KV, EM, KT

### Competing interests
All authors declare they have no competing interests.

### Data and materials availability
The full scripts and experiments are publicly available at https://github.com/sebnieh/QA_for_WhiteMatterLesionSegmentation. All data used are publicly available at the cited references.men

## Figures and Tables

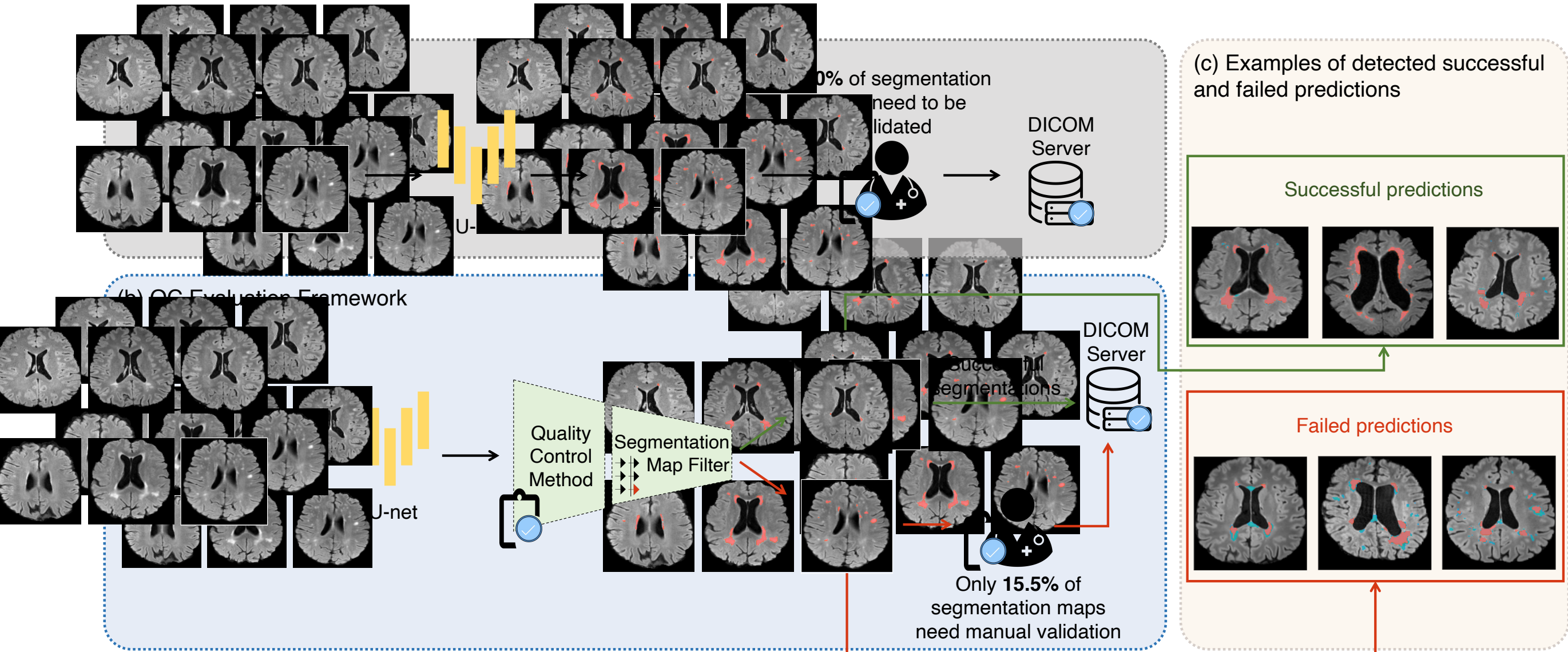

**Figure 1.** Comparison of (a) a clinical workflow and (b) QC evaluation framework and (c) examples of successful and failed predictions. In the clinical workflow, a segmentation model is applied to the MRI scans and outputs predicted WMH. The trained medical professional validates the quality of each prediction, manually delineates the failed segmentations and transfers them into DICOM server. The QC evaluation framework includes the application of a QC method and a segmentation map filter which identifies failed segmentation maps and sends them for manual segmentation. As demonstrated in (c), failed prediction maps are oftentimes undersegmented and require manual delineation. The approach shown in (b) optimises the time spent by clinician to only review the detected failed cases which on average happens 15% of the time based on our results.

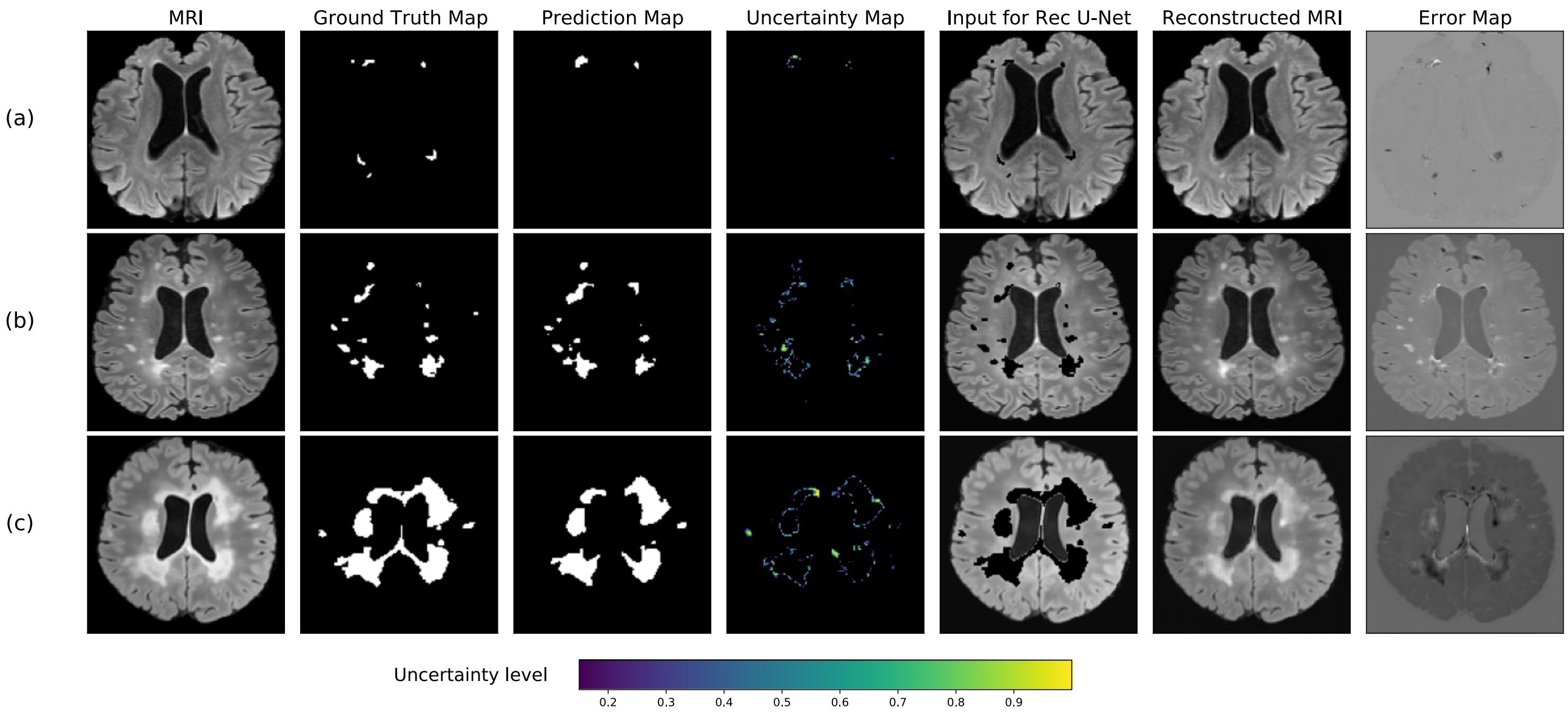

**Figure 2.** This figure provides examples of T1-weighted brain MRI, segmentation ground truth map, prediction map, uncertainty map, input for Rec U-net model, prediction of Rec U-net and resulted difference, namely error map. The examples are given for MRI volumes with (a) small WMH, (b) medium WMH and (c) large WMH load. We note that the model sometimes undersegments in (a) as well as the in (c). We also see that the uncertainty is mostly located on the borders of the WMH regions and that the difference is more present on the error maps for large WMH.

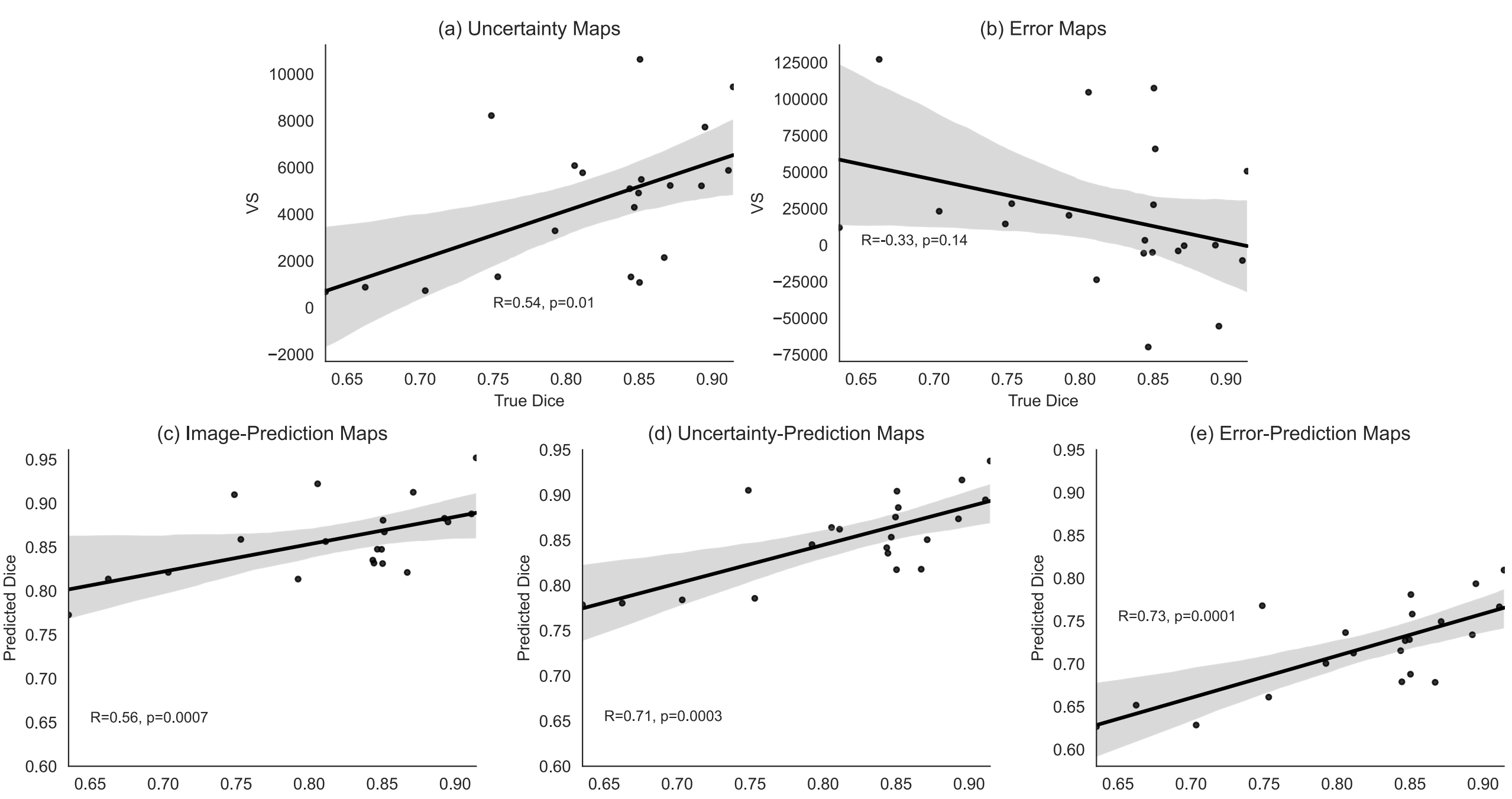

**Figure 3.** Scatterplots with a regression line of true Dice values and QC measurements: aggregated VS values from (a) uncertainty and (b) error maps and predicted Dice values based on input of (c) image-prediction pair, (d) uncertainty-prediction pair and (e) error-prediction pair. We can see in (a) a linear positive relationship between the values of uncertainty VS and Dice. In (b), we note that the correlation is insignificant (R=-0.33, p <0.14). In (e), we note that the predicted Dice values are slightly more correlated with true Dice values than in (c-d). The illustrated results are given for fold 1.

**Table 1.** QC performance results of VS aggregation and Dice Regression experiments with different input data. The mean results are reported, median is given in the brackets. Mean Dice before removing failed segmentation results is 0.84 (0.836).

| | **VS Aggregation Performance** | | **Dice Regression Performance** | | |
|---|---|---|---|---|---|
| Input | Uncertainty | Error | Image - Prediction | Uncertainty - Prediction | Error - Prediction |
| *Correlation coefficient* | 0.52 (0.548) | -0.2517 (-0.327) | 0.572 (0.566) | 0.712 (0.7108) | 0.677 (0.733) |
| *Dice after filtering* | 0.8504 (0.8496) | 0.837 (0.8307) | 0.848 (0.847) | 0.852 (0.8467) | 0.860 (0.8708) |
| *Precision* | 0.929 (0.937) | 0.477 (0.437) | 0.905 (0.895) | 0.968 (1.0) | 0.826 (0.941) |
| *Recall* | 0.887 (0.895) | 0.829 (0.778) | 0.837 (0.809) | 0.867 (0.809) | 0.868 (0.833) |
| *N failed segmentations identified* | 3.2 (3.0) | 9.8 (9.0) | 2.6 (2.0) | 1.8 (0.0) | 4.8 (3.0) |
| *MAE* | - | - | 0.0537 (0.0535) | 0.0486 (0.045) | 0.059 (0.057) |

**Table 2.** Characteristics of the MR image databases used in the experiments.

| **Dataset** | **Location** | **Scanner Name** | **Voxel Size (mm3) of of T1 & FLAIR** | **Size of T1 & FLAIR Scans** | **Sample Size** | **Demographics** | |
|---|---|---|---|---|---|---|---|
| | | | | | | **Male to female ratio** | **Age** |
| MICCAI17 | The Netherlands | 3T Philips Achieva | 0.96×0.95×3.00 T1 & FLAIR | 240×240×48 | 20 | Not given | Not given |

| | Singapore | 3T Siemens TrioTim | 1.00×1.00×3.00 T1 & FLAIR | 232×256×48 | 20 | | |
|---|---|---|---|---|---|---|---|
| | The Netherlands | 3T GE Signa HDxt | 0.98×0.98×1.20 T1 & FLAIR | 132×256×83 | 20 | | |
| MS15 | The Netherlands | 3T Philips Tesla | 1.00×1.00×1.00 T1 & FLAIR | 181×217×181 | 5 | 0.25 | 43.5±10.3 |
| MS16 | France | 3T Philips Ingenia | 0.74×0.74×0.85 T1w<br>0.74×0.74×0.7 FLAIR | 336×336×261 | 5 | 0.87 | 45.5 ± 7.8 |
| | | 3T Siemens Verio | 1×1×1 T1w<br>0.5×0.5×0.11 FLAIR | 256×256×176 T1w<br>512×512×144 FLAIR | 5 | 0.36 | 43.6±12.6 |
| MS17 | Slovenia | 3T Siemens Magnetom Trio | 0.42×0.42×3.30 T1w<br>0.47×0.47×0.80 FLAIR | 192x512x512 | 30 | 0.31 | Median: 39.1 |

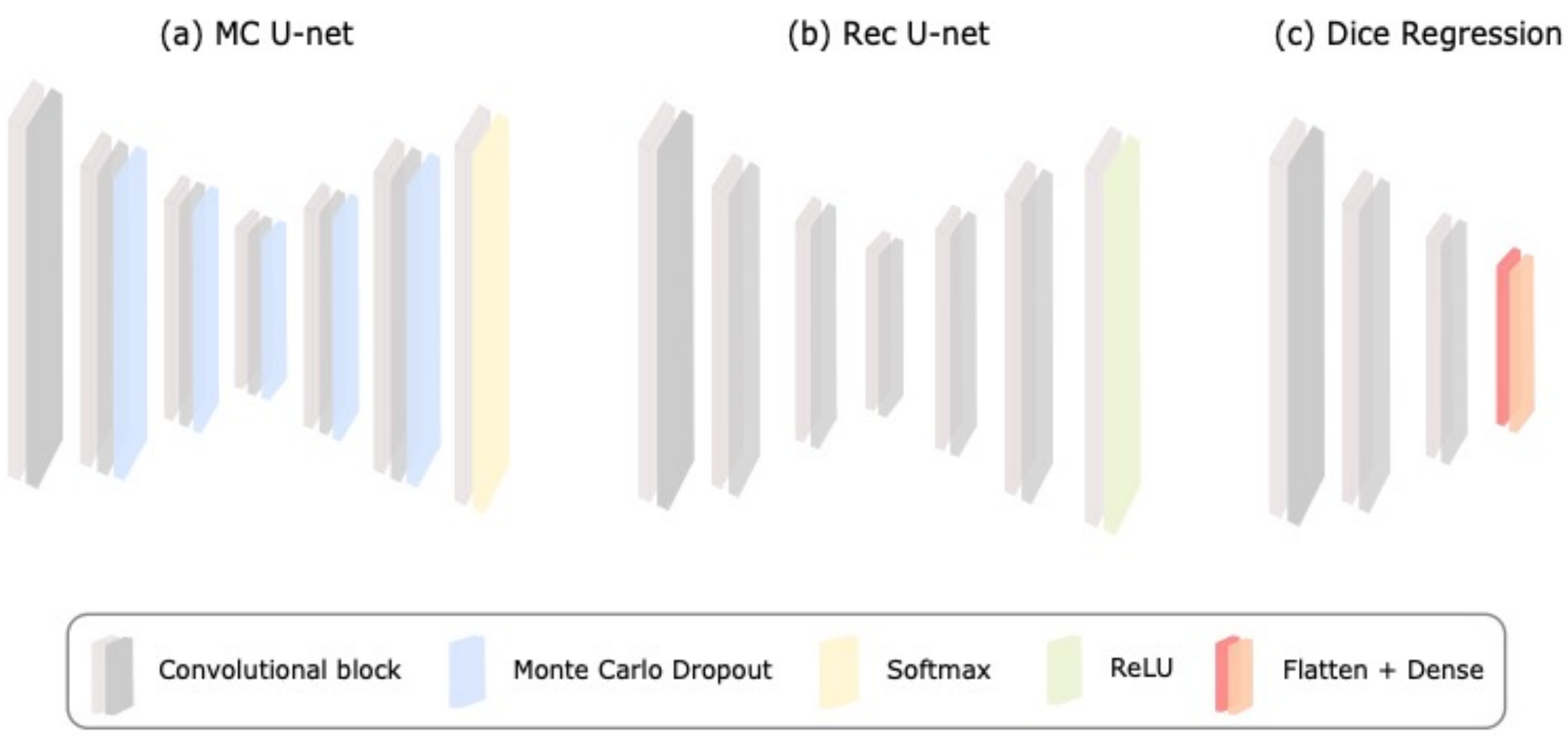

**Figure 4.** This figure illustrates the architecture of 3 networks used in the experiments: (a) MC U-net, (b) Rec U-net and (c) Dice Regression. In (a) we can see an illustration of modified U-net architecture with MC dropout layers. In (b) the U-net architecture was adjusted for a reconstruction task. In (c) we present the architecture of 3D CNN regression, namely Dice Regression, with features of size 16, 32, 64, 128 and linear activation function in the final Dense layer.